# Time-variant Seismic Resilience Analysis Model for Water Distribution Systems

Weinan Li[*,1], Ram K. Mazumder[2], and Yue Li[3]

[1] *Department of Civil and Environmental Engineering, Case Western Reserve University, USA*
[2] *Dept. of Civil, Environmental and Architectural Eng., University of Kansas, USA*
[3] *Department of Civil and Environmental Engineering, Case Western Reserve University, USA*

**Abstract**

Water distribution systems (WDS) sustained severe damage in the past earthquakes. While previous studies investigated the seismic performance of buried water pipelines, the effects of corrosion on the pipeline's seismic performance were ignored. The presence of corrosion on metallic pipeline walls aggravates seismic damage level because corrosion significantly reduces the pipeline's strength. Most of the existing buried pipelines in the United States are aged and non-ductile metallic pipelines (e.g., cast iron), which are vulnerable to seismic loading. To ensure continuous and smooth water supply to communities during and after earthquakes, it is necessary to evaluate the system-level seismic performance of WDS considering the aging effect in corroded pipelines. The current study develops a new framework of estimating the seismic resilience of WDS considering the time-variant effect of corrosion. The study formulates an approach that : (1) determines the seismic failure probability of pipeline using an extended American Lifelines Alliance (ALA) model that account for the effects of time-dependent corrosion; and (2) estimates system-level seismic performance based on pipeline's reliability and edge betweenness centrality. The proposed approach is illustrated with a scenario earthquake hazard for mid-size WDS. The outcomes of the study reveal that the presence of corrosion on pipelines significantly reduces the system-level seismic performance of WDS. Most cast iron pipes have 100 years lifetime, system-level seismic resilience may decrease by 81% at high seismic wave intensity.

## 1. Introduction

Water distribution systems (WDS) play an important role in economic and social wellbeing of communities. However, WDS sustained severe damages in past earthquakes, especially for aged water pipelines in regions of high seismicity. For instance, 2011 Christchurch earthquake resulted in 2311 leaks and breaks in water pipelines [1] and 1995 Hyogoken-Nambu earthquake damaged 1610 buried water pipelines [2]. Moreover, existence of corrosion deterioration on water pipeline wall amplifies the seismic damage likelihood and increases the susceptibility of system-level failure during an earthquake [3]. In the United States, a majority of existing buried water pipelines are made of cast iron, which experienced significant deterioration and broke over the last decades, and are still running after their expected design life [4]. Hence, seismic performance estimation considering corrosion effect is crucial for existing WDSs in their asset management plans.

To measure system-level performance after extreme events, Bruneau et al. [5] proposed the concept of resilience that is conceptualized as consisting of four dimensions: robustness, resourcefulness, redundancy, and rapidity. The seismic resilience metric of WDS is used to measure residual functionality of the system after earthquakes. Previous researchers have put efforts to study seismic impacts on WDSs by proposing resilience metrics. Todini [6] developed an energy-based metric to calculate the intrinsic capacity to overcome sudden failures of pipelines for WDS. Prasad and Park [7] modified Todini's resilience metric that took surplus energy and reliable system loops into consideration. Farahmandfar et al. [8] proposed a topology-based resilience metric for measuring connectivity condition of WDS. Additionally, Farahmandfar and Piratla [9] developed an energy-based resilience metric, and then compared the effectiveness between topology-based and energy-based resilience metrics. Balaei et al. [10] proposed a robustness metric based on vulnerability, redundancy, and criticality aspects of WDS.

However, none of previous mentioned studies considered corrosion effect in their resilience analysis. It was observed that aged pipelines experienced more severe damage compared to newly installed pipelines [11], and thus time-variant corrosion effect need to be considered in system-level seismic resilience analysis. To consider corrosion impacts, this study calculates seismic failure probability of pipelines by extending American Lifelines Alliance (ALA) fragility function with a strength reduction modifier. Then system-level seismic resilience is estimated using the existing resilience metric and a novel topology-based resilience metric.

## 2. Time-dependent Pipeline Failure Probability

The seismic resilience of WDS is evaluated using a scenario-based earthquake hazard in which seismic intensity (e.g., Peak Ground Velocity, Permanent Ground Displacement). The advantages of scenario-based seismic resilience analysis are that they can avoid practical difficulties in seismic study for WDS, and results obtained from scenario-based seismic study can be easily interpreted [3]. Since water pipes are buried underground and a large part of WDS is typically susceptible to seismic wave propagation during an earthquake, PGV is considered to compute pipeline seismic failure probability of pipelines.

Pipeline seismic failure probability is calculated using the guideline developed by ALA [12]. ALA [12] used 'repair rate' per 1000 ft pipeline length to represent pipeline vulnerability function that was developed based on past earthquake data sets, as shown in Eq. (1). Once repair rate is calculated, pipeline failure probability, assumed as a Poisson distribution of repair rate, is calculated using Eq. (2) [12]:

$$rr = K_1 \times (0.00187) \times PGV \quad (1)$$

$$P_f = 1 - e^{-rr \times L} \quad (2)$$

where $rr$ is the number of repairs per 1000 feet pipeline length, $L$ is pipeline length, $K_1$ is the modification factor, and $K_1$ values can be obtained from [12].

---

[*] E-Mail: wxl556@case.edu

However, this model (i.e., Eq. (2)) ignored corrosion impact. To take corrosion impact into consideration, Eq. (2) need to be modified. Since the maximum stress on pipeline increases due to corrosion growth, the impact of time-variant corrosion can be estimated by determining time-variant stress. Since corrosion and earthquakes have distinct impacts, it is practical to estimate these two failure effects separately, and then combine them to obtain time-dependent pipeline fragility function. Mazumder et al. [3] modified Eq. (2) by adding a strength reduction modifier (SRF). SRF is defined as the ratio of stress at a certain time considering corrosion impact to stress at uncorroded state, as shown in Eq. (3):

$$SRM(t) = \frac{\sigma(t)}{\sigma_0} \quad (3)$$

where $\sigma(t)$ is pipeline stress at time $t$ considering corrosion impact, $\sigma_0$ is the stress for uncorroded pipeline. The modified pipeline seismic failure probability function considering aging effect is expressed as follows [3]:

$$P_f(t) = 1 - e^{-SRM(t) \times rr \times L} \quad (4)$$

Because Eq. (3) is only the definition SRM, a specific equation is required so that it is computable practical. Ji et al. [13] performed a series of finite element analysis to obtain time-dependent stress enhancement due to corrosion, and then they derived a non-linear regression model to estimate strength reduction modifier. The regressed function for calculating strength reduction modifier is expressed as follows:

$$SRM(t) = \frac{\left\{1 - \alpha_1 \left[1 - \left(\frac{d'(t)}{d}\right)^{\alpha_2}\right] \times M(\Omega, R, d', v)\right\}}{\left\{1 - \alpha_1 \left[1 - \left(\frac{d'(t)}{d}\right)^{\alpha_2}\right]\right\}} \quad (5)$$

$$M = \frac{\alpha_3 \times \left[\frac{\sqrt[4]{3(1-v^2)}}{2} \cdot \left(\frac{\Omega}{\sqrt{Rd}}\right)\right]^{\alpha_4} \times \left(\frac{d'(t)}{d}\right)^{\alpha_5} + \alpha_6 \times \left(\frac{\Omega}{R}\right)^{\alpha_7}}{\alpha_8 \times \left(\frac{\Omega}{d'(t)}\right)^{\alpha_9} + \left(\frac{d'(t)}{d}\right)^{\alpha_{10}}}$$

(6)

where $d$ is pipeline uncorroded thickness, $d'(t)$ is remaining thickness after corrosion, $\Omega$ is corrosion pit radius, $R$ is pipeline radius, $v$ is Poisson ratio, and $\alpha_1$ to $\alpha_{10}$ are coefficients. The model coefficients $\alpha$ ($\alpha_1$: 0.9598, $\alpha_2$: 6.3792, $\alpha_3$: -0.0391, $\alpha_4$: 1.8741, $\alpha_5$: -1.1103, $\alpha_6$: 1.9858, $\alpha_7$: 0.0276, $\alpha_8$: 0.8762, $\alpha_9$: 0.0853, $\alpha_{10}$: 0.0762) are obtained from [13].

## 3. Seismic Resilience Metrics

The resilience metric of WDS is defined to calculate the residual functionality of the system after extreme events. Topological resilience metrics can be used to determine system's residual connectivity status after earthquakes. The present study uses topological resilience metric to determine system seismic performance because the connectivity is the basic requirement for network infrastructure systems. There is no universally accepted seismic resilience metric; hence, researcher often select performance indicators based on expert opinions and engineering judgements. One existing metric and a proposed modified metric are discussed herein.

### 3.1 Topology-based Resilience Metric (TBRM)

Farahmandfar et al. [8] proposed a topology-based resilience metric (TBRM) that was developed by considering pipelines reliability and modified nodal degree. In their metric, nodal degree refers to sum of connected pipelines' reliabilities, and water demand of the node is used as weight to represent to importance of the node. The time-variant TBRM is expressed as follows:

$$TBR(t) = \frac{\sum_{i=1}^{N_n}\{[\sum_{l=1}^{N_i}(1 - P_f(t))] \times Q_i\}}{4 \times \sum_{i=1}^{N_n} Q_i} \quad (7)$$

where $N_n$ is number of nodes in WDS, $N_i$ is number of connected pipelines to node $i$, $Q_i$ is the water demand of node $i$.

### 3.2 Modified Topology-based Resilience Metric (MTBRM)

There exists a problem with TBRM due to the coefficient "4" in its denominator. Farahmandfar et al. [8] developed TBRM based on an assumption that all $N_i$ are set as four. However, the nodal degree, defined as the number of pipes connected to a specific node, varies significantly corresponding to the different configurations of the WDSs, and thus TBRM cannot accurately predict topological resilience. This study modified TBRM as follows:

$$MTBR(t) = \frac{\sum_{i=1}^{N_n}\{[\sum_{l=1}^{N_i}(1 - P_f(t))] \times Q_i\}}{\sum_{i=1}^{N_n} N_i \times Q_i} \quad (8)$$

### 3.3 EBC-based Resilience Metric (EBRM)

Previous mentioned metric uses nodal water demand as weight to represent nodal importance; however, nodal water demand is a hydraulic characteristic. To develop a topological resilience metric that is not affected by hydraulic working condition, performance indicators of the resilience metric should be independent from hydraulic condition. Therefore, a novel topological resilience metric is proposed herein based on pipeline reliability and edge betweenness centrality (EBC) of pipeline. EBC of an edge is defined as the sum of the ratio of shortest paths between all possible pairs of nodes that pass the edge. Therefore, EBC of a pipeline can provide importance of the pipeline because failure of the component with greater EBC will definitely cause more severe water flow interruption. EBC of a pipeline it is defined as follows [14]:

$$C_l^B = \sum_{\substack{p \neq q \in V \\ l \in E}} \frac{\sigma_{u,o}(l)}{\sigma_{u,o}}  \qquad (9)$$

where $C_l^B$ is the edge betweenness centrality of pipeline $l$, $\sigma_{u,o}(l)$ is number of shortest paths between node $u$ to node $o$ that pass edge $l$, $\sigma_{u,o}$ is total number of shortest paths between node $u$ to node $o$, $V$ and $E$ refer to node sets and edge sets in WDS, respectively. The EBC-based topological resilience metric is expressed as follows:

$$EBRM(t) = \frac{\sum_{i=1}^{N_m}(1 - P_f(t)) \times C_i^B}{\sum_{i=1}^{N_m} C_i^B} \qquad (10)$$

where $N_m$ is total number of pipelines in WDS.

## 4. Case Study

Any-town, a virtual WDS designed by Walski et al. [15] is utilized to illustrate proposed framework, as shown in Fig. 1. The WDS is comprised of 22 nodes and 41 pipelines. The numbers next to the nodes and pipelines show node and pipe identification numbers, respectively. The numbers provided in the brackets show pipeline length in kilometers. All of required data (e.g., pipeline length, pipeline diameter, nodal water demand) for Any-town WDS are obtained from the University of Exeter Centre for Water Systems [3].

This study investigates seismic performance alternation trend of WDS considering two factors: (1) PGV and (2) Corrosion. PGV impact is considered by calculating the seismic resilience metrics as PGV increases. Corrosion impact is considered by computing time-dependent failure probability of pipeline using Eqs. (4) to (6), and then time-dependent seismic resilience metrics are calculated accordingly. Since majority of old water pipelines in the U.S. are made of cast iron, it is assumed that all pipelines in the illustrative Any-town WDS are cast iron pipelines. The corrosion growth rate is set as 0.12 mm/year according to empirical dataset of aged cast iron pipelines provided by Petersen and Melchers [16]. The initial corrosion pit is assumed to be zero, and corrosion pit radius is assumed to be five times of corrosion pit [16]. Earthquake attenuation model and spatial correlation are ignored in the present study for simplicity. In other words, all pipelines are subjected to same PGV value at same time. The proposed framework can be improved by considering earthquake attenuation and spatial correlation in following research.

Figs. (2) to (4) show effects of PGV and corrosion on TBRM, MTBRM, and EBRM, respectively. Resilience curves shown in part (a) of the three figures are created by varying PGV at six different ages of Any-town WDS. In contrast, resilience curves shown in part (b) of the three figures are generated by increasing exposure time of Any-town WDS subjected to 5 different PGV values. These three seismic metrics provide close outcomes, TBRM provides greatest and EBRM provides smallest values in comparison. For example, at PGV of 12 cm/s, TBRM equals to 0.741, 0.692, 0.514, 0.256, and 0.102 at 0, 30, 60, 90, and 120 years elapsed time, respectively. At 12 cm/s PGV value, MTBRM equals to 0.682, 0.637, 0.473, 0.235, and 0.094 at 0, 30, 60, 90, and 120 years elapsed time, respectively. At 12 cm/s PGV value, EBRM equals to 0.660, 0.614, 0.447, 0.216, and 0.087 at 0, 30, 60, 90, and 120 years of elapsed time, respectively. Similar outcomes can be obtained from other cases (e.g. various PGV values and elapsed times). Since these three metrics provide close outcomes, therefore, all of these three resilience metrics can be used to predict seismic performance of Any-town WDS.

Corrosion on pipeline wall significantly reduces WDS seismic performance. For example, at PGV of 12cm/s, MTBRM equals to 0.637, 0.473, 0.235, and 0.094 at 30, 60, 90, and 120 years of age, respectively, as shown in Fig. 3(b). At the same PGV value, EBRM equals to 0.614, 0.447, 0.216, and 0.087 at 30, 60, 90, and 120 years of exposure time, respectively, as shown in Fig. 4(b). Moreover, as shown in part (a) of the figures, seismic resilience decreases more quickly for aged system. System seismic resilience drops gradually as exposure time increases because pipeline failure probability increases over time, as shown in Eq. (4). Because cast iron pipes have typical life between 80 to 100 years [4], EBRM decreases by 32%, 53%, 66%, 75%, and 81% from uncorroded state to 100 years exposure time at PGV of 3 cm/s, 6 cm/s, 9 cm/s, 12 cm/s, and 15 cm/s, respectively. Similar outcomes are obtained for TBRM and MTBRM. Hence, seismic resilience of corroded pipe drops significantly, and greater PGV values amplify the resilience decrease.

PGV also has significant influences on seismic performance. As an example, at 90 years age, EBRM equals to 0.661, 0.447, 0.308, 0.216, and 0.154 at PGV of 3 cm/s, 6 cm/s, 9 cm/s, 12 cm/s, and 15 cm/s, respectively. It is because pipeline failure probability increases as PGV increases, as shown in Eqs. (1) and (4). EBRM decreases by 34%, 39%, 55%, 78%, 91%, and 99% at 0, 30, 60, 90, 120, and 150 years elapsed time, respectively. Similar results were obtained from TBRM and MTBRM. Older pipes are the most susceptible ones to seismic wave propagation.

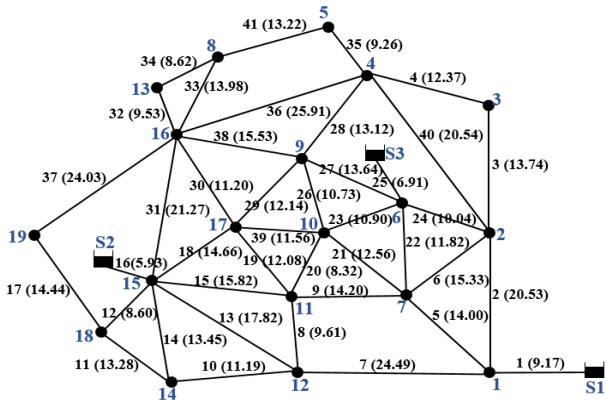

**Figure 1.** Layout of Any-town WDS

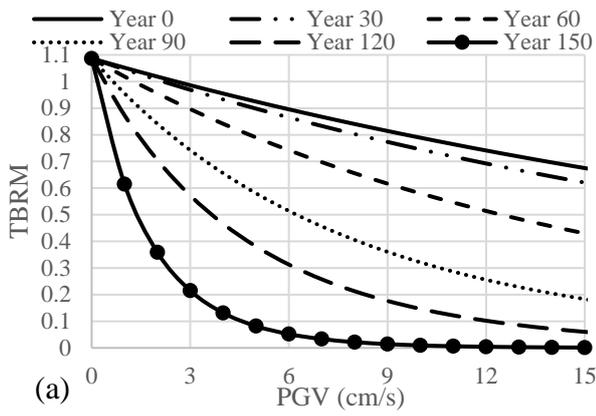

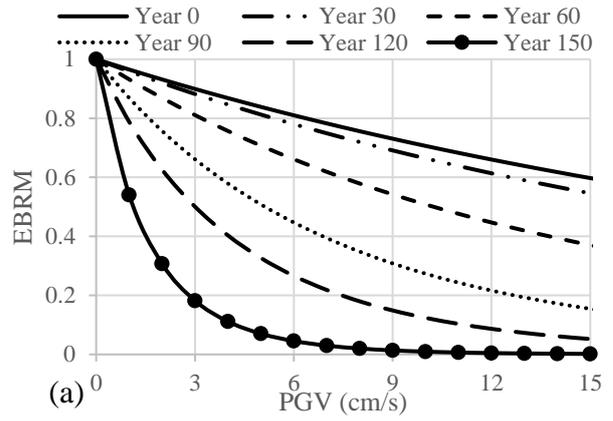

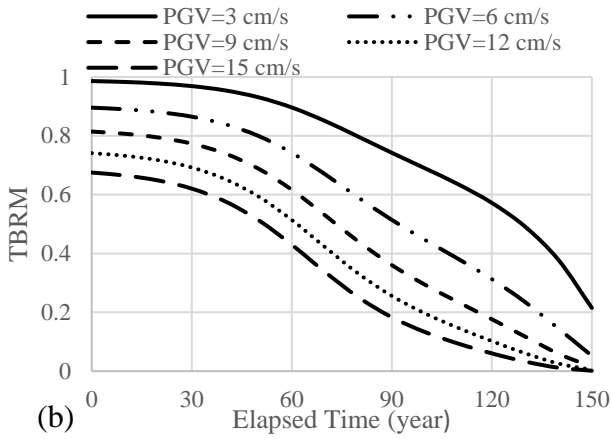

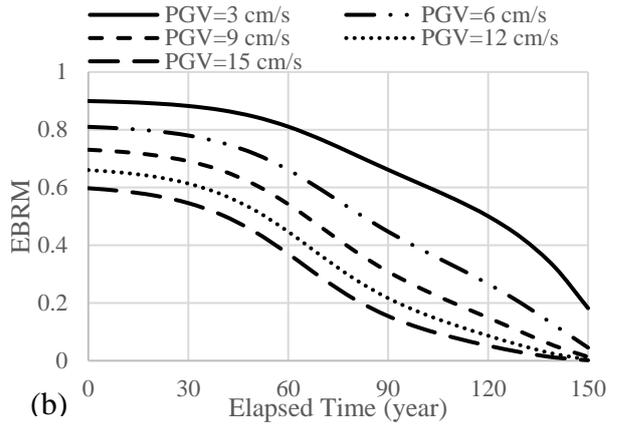

**Figure 2.** TBRM for Any-town WDS

**Figure 4.** EBRM for Any-town WDS

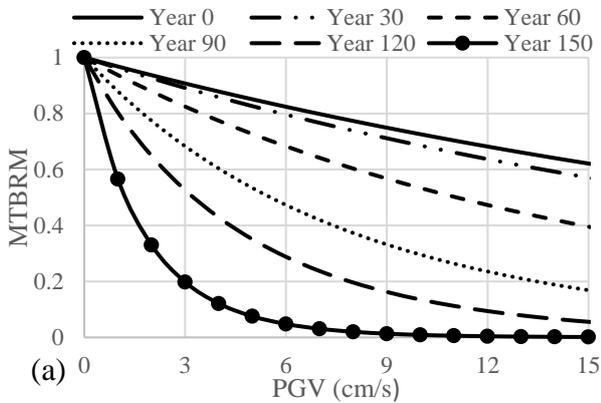

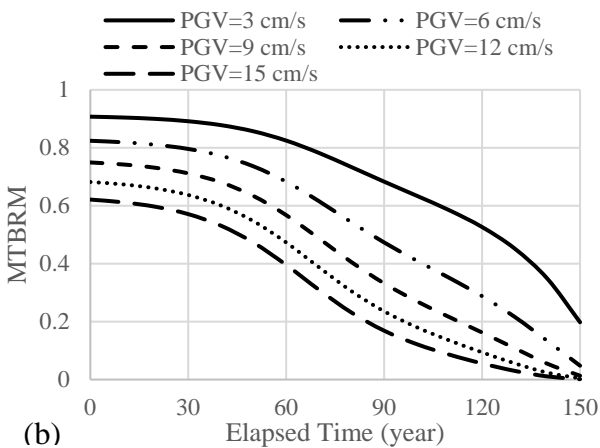

**Figure 3.** MTBRM for Any-town WDS

## 5. Conclusion

Large earthquakes can cause severe damage to aging water pipelines. It is required to estimate system residual functionality considering the impact of corrosion to develop an effective seismic risk management plan for WDS. The system-level seismic performance is represented by one existing topology-based resilience metric and a resilience metric. The corrosion effects are considered by adding a strength reduction modifier to the pipeline fragility function. Based on time-dependent pipeline failure probability, time-dependent system-level resilience can be obtained. Any-town WDS is utilized as a sample system to illustrate the proposed framework. The results revealed that TBRM, MTBRM, and EBRM provide similar resilience values at all conditions. Therefore, the three metrics are effective measures for determining system connectivity condition after earthquakes. System-level resilience decreases up to 81% at 100 years of age, it is concluded that the presence of corrosion has significantly negative impacts on seismic resilience of WDS.

**Acknowledgements**

The research described in this paper was supported, in part, by the National Science Foundation (NSF) Critical Resilient Interdependent Infrastructure Systems and Processes (CRISP) under Grant No. NSF-1638320. This support is thankfully acknowledged. However, the writers take sole responsibility for the views expressed in this